\documentstyle[a4,epsfig,11pt]{article}
\begin{document}

\baselineskip=7mm
\def\ap#1#2#3{           {\it Ann. Phys. (NY) }{\bf #1} (#2) #3}
\def\arnps#1#2#3{        {\it Ann. Rev. Nucl. Part. Sci. }{\bf #1} (#2) #3}
\def\cnpp#1#2#3{        {\it Comm. Nucl. Part. Phys. }{\bf #1} (#2) #3}
\def\apj#1#2#3{          {\it Astrophys. J. }{\bf #1} (#2) #3}
\def\asr#1#2#3{          {\it Astrophys. Space Rev. }{\bf #1} (#2) #3}
\def\ass#1#2#3{          {\it Astrophys. Space Sci. }{\bf #1} (#2) #3}

\def\apjl#1#2#3{         {\it Astrophys. J. Lett. }{\bf #1} (#2) #3}
\def\ass#1#2#3{          {\it Astrophys. Space Sci. }{\bf #1} (#2) #3}
\def\jel#1#2#3{         {\it Journal Europhys. Lett. }{\bf #1} (#2) #3}

\def\ib#1#2#3{           {\it ibid. }{\bf #1} (#2) #3}
\def\nat#1#2#3{          {\it Nature }{\bf #1} (#2) #3}
\def\nps#1#2#3{          {\it Nucl. Phys. B (Proc. Suppl.) } {\bf #1} (#2) #3}
\def\np#1#2#3{           {\it Nucl. Phys. }{\bf #1} (#2) #3}

\def\pl#1#2#3{           {\it Phys. Lett. }{\bf #1} (#2) #3}
\def\pr#1#2#3{           {\it Phys. Rev. }{\bf #1} (#2) #3}
\def\prep#1#2#3{         {\it Phys. Rep. }{\bf #1} (#2) #3}
\def\prl#1#2#3{          {\it Phys. Rev. Lett. }{\bf #1} (#2) #3}
\def\pw#1#2#3{          {\it Particle World }{\bf #1} (#2) #3}
\def\ptp#1#2#3{          {\it Prog. Theor. Phys. }{\bf #1} (#2) #3}
\def\jppnp#1#2#3{         {\it J. Prog. Part. Nucl. Phys. }{\bf #1} (#2) #3}

\def\rpp#1#2#3{         {\it Rep. on Prog. in Phys. }{\bf #1} (#2) #3}
\def\ptps#1#2#3{         {\it Prog. Theor. Phys. Suppl. }{\bf #1} (#2) #3}
\def\rmp#1#2#3{          {\it Rev. Mod. Phys. }{\bf #1} (#2) #3}
\def\zp#1#2#3{           {\it Zeit. fur Physik }{\bf #1} (#2) #3}
\def\fp#1#2#3{           {\it Fortschr. Phys. }{\bf #1} (#2) #3}
\def\Zp#1#2#3{           {\it Z. Physik }{\bf #1} (#2) #3}
\def\Sci#1#2#3{          {\it Science }{\bf #1} (#2) #3}

\def\n.c.#1#2#3{         {\it Nuovo Cim. }{\bf #1} (#2) #3}
\def\r.n.c.#1#2#3{       {\it Riv. del Nuovo Cim. }{\bf #1} (#2) #3}
\def\sjnp#1#2#3{         {\it Sov. J. Nucl. Phys. }{\bf #1} (#2) #3}
\def\yf#1#2#3{           {\it Yad. Fiz. }{\bf #1} (#2) #3}
\def\zetf#1#2#3{         {\it Z. Eksp. Teor. Fiz. }{\bf #1} (#2) #3}
\def\zetfpr#1#2#3{         {\it Z. Eksp. Teor. Fiz. Pisma. Red. }{\bf #1} (19#2) #3}
\def\jetp#1#2#3{         {\it JETP }{\bf #1} (19#2) #3}
\def\mpl#1#2#3{          {\it Mod. Phys. Lett. }{\bf #1} (#2) #3}
\def\ufn#1#2#3{          {\it Usp. Fiz. Naut. }{\bf #1} (#2) #3}
\def\sp#1#2#3{           {\it Sov. Phys.-Usp.}{\bf #1} (#2) #3}
\def\ppnp#1#2#3{           {\it Prog. Part. Nucl. Phys. }{\bf #1} (#2) #3}
\def\cnpp#1#2#3{           {\it Comm. Nucl. Part. Phys. }{\bf #1} (#2) #3}
\def\ijmp#1#2#3{           {\it Int. J. Mod. Phys. }{\bf #1} (#2) #3}
\def\ic#1#2#3{           {\it Investigaci\'on y Ciencia }{\bf #1} (#2) #3}
\def\tp{these proceedings}
\def\pc{private communication}
\def\ip{in preparation}
\newcommand{\TeV}{\,{\rm TeV}}
\newcommand{\GeV}{\,{\rm GeV}}
\newcommand{\MeV}{\,{\rm MeV}}
\newcommand{\keV}{\,{\rm keV}}
\newcommand{\eV}{\,{\rm eV}}
\newcommand{\Tr}{{\rm Tr}\!}
\renewcommand{\arraystretch}{1.2}
\newcommand{\be}{\begin{equation}}
\newcommand{\ee}{\end{equation}}
\newcommand{\bea}{\begin{eqnarray}}
\newcommand{\eea}{\end{eqnarray}}
\newcommand{\ba}{\begin{array}}
\newcommand{\ea}{\end{array}}
\newcommand{\bmat}{\left(\ba}
\newcommand{\emat}{\ea\right)}
\newcommand{\refs}[1]{(\ref{#1})}
\newcommand{\ler}{\stackrel{\scriptstyle <}{\scriptstyle\sim}}
\newcommand{\ger}{\stackrel{\scriptstyle >}{\scriptstyle\sim}}
\newcommand{\lag}{\langle}
\newcommand{\rag}{\rangle}
\newcommand{\ns}{\normalsize}
\newcommand{\cm}{{\cal M}}
\newcommand{\gr}{m_{3/2}}
\newcommand{\p}{\partial}
\renewcommand{\le}{\left(}
\newcommand{\ri}{\right)}
\renewcommand{\o}{\overline}

\relax
\def\321{$SU(3)\times SU(2)\times U(1)$}
\def\21{$SU(2)\times U(1)$}
\def\dbd{$0\nu\beta\beta~$}
\def\ord{{\cal O}}
\def\mnu0{{\cal M}_{0\nu}}
\def\dt{\delta_{\tau}}
\def\tl{{\tilde{l}}}
\def\tL{{\tilde{L}}}
\def\bd{{\overline{d}}}
\def\tL{{\tilde{L}}}
\def\a{\alpha}
\def\b{\beta}
\def\g{\gamma}
\def\c{\chi}
\def\d{\delta}
\def\D{\Delta}
\def\db{{\overline{\delta}}}
\def\Db{{\overline{\Delta}}}
\def\e{\epsilon}
\def\l{\lambda}
\def\n{\nu}
\def\m{\mu}
\def\nt{{\tilde{\nu}}}
\def\p{\phi}
\def\P{\Phi}
\def\solm{\Delta_{S}}
\def\sola{\theta_{S}}
\def\mee{m_{ee}}
\def\atm{\Delta_{\makebox{\tiny{\bf atm}}}}
\def\k{\kappa}
\def\x{\xi}
\def\r{\rho}
\def\s{\sigma}
\def\t{\tau}
\def\th{\theta}
\def\om{\omega}
\def\ne{\nu_e}
\def\nm{\nu_{\mu}}
\def\snui{\tilde{\nu_i}}
\def\ehat{\hat{e}}
\def\la{{\makebox{\tiny{\bf loop}}}}
\def\ta{\tilde{a}}
\def\tb{\tilde{b}}
\def\mb{m_{1b}}
\def\mt{m_{1 \tau}}
\def\rl{{\rho}_l}
\def\meg{\m \rightarrow e \g}
\renewcommand{\Large}{\large} \title{\hfill hep-ph/0302181\\[1.0cm]
\large \bf Implications of partially
degenerate neutrinos at a high scale in the light of KamLAND and WMAP} \author{ Anjan S. Joshipura and
Subhendra Mohanty\\[.5cm] {\ns\it Theoretical Physics Group, Physical
Research Laboratory,}\\ {\ns\it Navrangpura, Ahmedabad 380 009, India.}}
\date{} \maketitle \vskip .5cm 
\begin{center} {\bf Abstract}\end{center} \vskip 1.0cm
Electroweak radiative corrections can generate the neutrino (mass)$^2$
difference required for the large mixing angle solution (LMA) to the solar
neutrino problem if two of the neutrinos are assumed degenerate at high energy.
We test this possibility with the existing experimental knowledge of the low
energy neutrino mass and mixing parameters. We derive restrictions on ranges of
the high scale mixing matrix elements and obtain predictions for the low energy
parameters required in order to get the LMA solution of the solar neutrino
problem picked out by KamLAND. We find that in the case of standard model this
is achieved only when the (degenerate) neutrino masses lie in the range
$(0.7-2) \eV$ which is at odds with the cosmological limit $m_{\nu}<0.23 \eV$
(at $95 \% C.L$)  established recently using WMAP results. Thus SM radiative
corrections cannot easily generate the LMA solution in this scenario. However,
the LMA solution is possible in case of the MSSM electroweak corrections with
(almost) degenerate spectrum or with inverted mass hierarchy for limited ranges
in the high scale parameters.
\newpage 
\noindent{\it Introduction:} Results from solar and atmospheric neutrinos
\cite{revs} have
greatly helped in establishing patterns for neutrino masses and mixings
particularly after the report of the positive evidence of neutrino
oscillations seen at KamLand \cite{kam}. The allowed possibilities are
quite constrained. One needs hierarchical differences in the neutrino
(mass)$^2$ and  two large and one small ($
\leq 0.2$) mixing angle to fit the observations.

The phenomenological determination of neutrino masses and mixing raises
several theoretical questions two of which are ($i$) why two of the six
physical fermionic mixing angles are large and ($ii$) what is the cause of
hierarchy in the solar ($\solm$) and the atmospheric ($\atm$) mass scales.
There have been number of answers to these questions in variety of
frameworks \cite{revs}. One possibility is to invoke radiative corrections
to understand smallness of ${\solm\over \atm}$. These corrections could be
weak corrections to the lepton number violating neutrino mass operators
\cite{wolf} or could also come from physics beyond standard model
\cite{emtsym,sf}. Possibility of the electroweak corrections generating
solar scale has recently been analyzed in
\cite{pdp,pdp1,pdp2}. It is
assumed that two of the neutrinos are degenerate at some high scale and
electroweak corrections result in generation of the solar (mass)$^2$
difference. This
possibility was shown to be quite constrained. It leads to a definite
prediction for the solar scale namely,  
\be\label{pred} \solm \cos
2\sola=4 \delta_{\tau} \sin^2\theta_A |m_{ee}|^2
+\ord(\delta_{\tau}^2)~.
\ee
Here $\solm$ is the mass-squared difference responsible for the solar neutrino
oscillations, and the angles $\sola$ and $\theta_A$ respectively denote
the solar
and the atmospheric mixing angles at a low scale. $m_{ee}$ is the
effective neutrino mass probed in the \dbd decay and
$\delta_\tau$ specifies the size of the radiative corrections induced
by the Yukawa coupling of the $\tau$:
\be \label{delta1}
\delta_\tau\approx c\left({m_\tau\over 4 \pi v }\right)^2 \ln{M_X\over M_Z}~.
\ee
$c=\frac{3}{2},-\frac{1}{\cos^2\b}$ in case of the standard
model (SM) and  the minimal supersymmetric standard model (MSSM)
respectively \cite{rg,radrev,rad}.

Additional assumption made in deriving eq.(\ref{pred}) was that the  
mixing
element $U_{e3}$ of the leptonic mixing matrix $U$ was zero at  high  
scale. This assumption was motivated
by the observed smallness of $U_{e3}$ at low energy. Given this
assumption, eq.(\ref{pred}) makes very strong predictions analyzed in
detail in \cite{pdp1,pdp2}. Two of the major consequences being that the
MSSM radiative corrections cannot generate the large mixing angle (LMA)
solution to the solar neutrino problem and in the case of SM one needs
$m_{ee}$ close to the present experimental limit \cite{limit}.

A non-zero and relatively large low scale $U_{e3}\sim
0.2$
could change some of the
qualitative aspects of predictions based on eq.(\ref{pred}). More
importantly, the mixing angle at high scale could even be larger than
$0.2$ and can still give rise to an experimentally acceptable $U_{e3}$ at the low 
scale. In this  paper we study the general implications of two degenerate
neutrinos at high scale without assuming a zero $U_{e3}$. The general
numerical analysis carried out in this paper leads to very strong
predictions for  three of the yet unknown observables
 namely, $U_{e3},m_{ee}$ and the  absolute neutrino mass 
$m_{\nu_{e}}$ probed in beta decay \cite{mainz}.
One of the conclusions of our analysis is that a non-zero high scale $U_{e3}$ makes the MSSM viable. In the SM however
the LMA mass difference  can be generated  radiatively only
when the  the degenerate neutrino mass is in the range of $(0.7-2)~ eV$. 
This requirement is ruled out by the WMAP result that
$\Omega_\nu h^2 < 0.0076$ (at $95 \% C.L$.) \cite{wmap}
which for the degenerate neutrino spectrum implies that $m_\nu < 0.23.$

\noindent {\it Formalism:} Consider a CP conserving theory specified by a
general $3\times 3$ real
symmetric neutrino mass matrix $M_{\nu 0}$ specified at a high scale
$M_X$. We require that the solar scale vanishes at $M_X$ and consequently
two of the eigenvalues of $M_{\nu 0}$ are degenerate, $i.e. $ , we assume
\footnote{The solar angle can be rotated away in case of the other viable
possibility $m_{\nu_{0i}}=(m,m,m')$ which cannot reproduce the observed
pattern.}  
$m_{\nu_{0i}}=(m,-m,m')$ for the neutrino masses at $M_X$. The atmospheric
neutrino oscillations are
induced by $\Delta_{A0}\equiv |m'^2-m^2|$.

Neutrino mixing matrix has the following general form under the assumption of 
CP  conservation.
\be \label{u0} U_0=R_{23}(\theta_2) R_{13}(\theta_3) R_{12}(\theta_1) ~,
\ee
where $R_{ij}(\theta)$ denotes a rotation in the $ij^{th}$ plane by 
an angle $\theta$. The $\theta_{1,2}$ are assumed to vary between 0 and 
$\pi/2$ while $s_3=\sin\theta_3$ varies over the full range. The
neutrino mass matrix 
at $M_X$ is given by
\be\label{zeroorder}
 \mnu0=U_0~ Diag.(m,-m,m')~ U_0^T ~. \ee
The matrix $\mnu0$ determined by eq.(\ref{zeroorder}) is modified
by the radiative corrections. The radiatively corrected form of
$\mnu0$ follows from the relevant RG equations \cite{rg}. We assume the RG
equations corresponding to the SM or the MSSM. The modified
neutrino mass matrix is given \cite{radrev} in this case by
\be \label{corrected} \mnu0\rightarrow {\cal M}_\nu \approx I_g
I_t ~(I~ U_0 ~diag.(m,-m,m')~U_0^T ~I~)\ee
where $I_{g,t}$ are calculable numbers depending on the gauge and
top quark Yukawa couplings. $I$ is a flavour dependent matrix
given by
$$ I\approx diag.(1+\delta_e,1+\delta_\mu,1+\delta_\tau) ~.$$
$\delta_{e,\mu}$ are obtained from  eq.(\ref{delta1}) by replacing
the tau mass by the electron and the muon masses respectively. The
physical neutrino masses and mixing are obtained by diagonalizing
the above matrix. 

The eigenvalues of $ {\cal M}_\nu $ can be approximately determined in the
limit of vanishing $m_{e,\mu}$. We find, 
\bea  \label{ev} m_{\nu_1}&\approx& m(1+2 \dt (c_1c_2s_3-s_1s_2 )^2) 
+\ord(\dt^2) ~, \nonumber \\ 
m_{\nu_2}&\approx&- m(1+2 \dt (s_1c_2s_3+c_1s_2  )^2) +\ord(\dt^2) ~,
\nonumber \\ 
m_{\nu_3}&\approx& m' (1+2 \dt c_2^2c_3^2)+\ord(\dt^2) ~. \eea
We thus have
\be \label{d21} \Delta_{21}\equiv m_{\nu_2}^2-m_{\nu_1}^2\approx 4 \dt m^2
\left( \cos 2 \theta_1 (s_2^2-s_3^2 c_2^2)+ 
s_3 \sin 2 \theta_1 \sin 2 \theta_2 \right)+\ord(\dt^2) ~. \ee

It is conventional to order masses in a way that makes the solar scale
$\solm$ positive.  The phenomenological analysis then restricts 
$\sola$ to be $<\pi/4$. The $\Delta_{21}$ as defined above can have
either sign.
For positive $\Delta_{21}$, $\solm=\Delta_{21}$ and $\cos2 \sola=\cos
2\theta_1$. One needs to interchange first two eigenvalues and eigenvector
of ${\cal M}_\nu$ in the
opposite case with $\Delta_{21}<0$ giving us $\solm=-\Delta_{21}$ and
$\cos2 \sola=-\cos 2\theta_1$. In either situation, one needs
$\Delta_{21}\cos2 \theta_1>0$ in order to obtain the LMA solution. For
$s_3=0$, the sign of $\dt$ determines the sign of $\Delta_{21}\cos2
\theta_1$. As a result one cannot reproduce the LMA solution in the case
of MSSM as already remarked \cite{pdp1,pdp2}. The introduction of a
non-zero $s_3$ changes
this behavior. As follows from eq.(\ref{d21}) one can now obtain positive
$\Delta_{21}\cos2 \theta_1$ in case of the MSSM also for some range in
parameter space corresponding to $s_3\cos 2\theta_1<0$. This range is
limited since $s_3$ cannot be very large without conflicting with the
observed bound on $U_{e3}$.

Realization of the LMA solution in this framework would require some
restrictions on the initial high scale parameters and would also restrict
the values of the low energy observables. These observables include
$m_{ee},\solm,U_{e3}$ and the electron neutrino mass $m_{\nu_i}$. 
We study these restrictions in detail below through the
following procedure:

\noindent {\it Input parameters:} We randomly vary the input mixing angles
$\theta_i$ and the mass $m$ over the range $0<\theta_{1,2}<\pi/2$,
$-\pi/2< \theta_3<\pi/2$ and $0<m<2 \eV$. Since the atmospheric mass scale does
not significantly change by the radiative corrections, we fix the third
mass $m'$ by requiring $\atm\equiv |m'^2-m^2|=(.05)^2$. $m'$ can have
either sign relative to $m$ and could be heavier or lighter than $m$. We
will consider $|m'|>|m|$ and $m'\ll m$. The MSSM radiative
corrections involve an additional parameter $\cos \b$ which is also
randomly varied in the range (0-1].

\noindent {\it Output values of observables:} Eq.(\ref{corrected}) is
numerically diagonalized for each choice of the randomly chosen input
variables. This gives us the output values of the solar and atmospheric
masses and mixing angles, $U_{e3}$ as well as $m_{ee}$ and
$m_{\nu_e}$. The known output parameters are required to lie in the
range \cite{revs,sola}:
\be \label{output}
\ba{cc}
~~~~~~~~~~~~~~~~~~~~~~~~~~~~~~~~~~~~~~~~~~|U_{e3}|\leq 0.2&\\
4\cdot 10^{-5}\eV^2\leq\solm\leq 2.8\cdot 10^{-4}\eV^2~~;&~~~~~
0.2\leq\tan^2\sola\leq 0.8 ~, \\
1.2\cdot 10^{-3}\eV^2\leq\atm\leq 5.0\cdot 10^{-3}\eV^2~~;&~~~~~
0.8\leq \sin^2 2\theta_A\leq 1.0 ~, \\ \ea
\ee
The ranges quoted for the solar parameters is $3\sigma$ level. The KamLand
results on the anti neutrino oscillations do not allow the
full range of the LMA solution quoted above but the allowed values 
at 90\% CL lie in
two different ranges which are  subsets of the above range. Consistency of
the results with KamLand can be explicitly seen from the figures to be
presented.

From the randomly varied set of 100,000 points, we collect the acceptable
choice of the (high scale) input variables which lead to the low energy
parameters lying
in the ranges in eq.(\ref{output}). This procedure also gives us the
predicted values of the output observables like $m_{ee},m_{\nu_e}$
corresponding to the acceptable choices of the input parameters. Results
of our analysis are presented in case of the SM as well as MSSM in
Figs.(1-5).

Fig.(1) shows the allowed values of the input parameters $s_1,s_3$
consistent with random variations of these and  other parameters. The
effect of the radiative corrections on mixing angles is not pronounced in
case of the SM and the input ranges for $s_3$ and $\tan^2\theta_1$
coincide approximately with the
allowed ranges in $U_{e3}$ and $\tan^2\sola$. Radiative corrections can be
appreciable in case of the MSSM and an $s_3$ as large as 0.4
can lead to $U_{e3}\leq 0.2$ at the low scale. But values of $s_3$ larger
than this cannot
reproduce the correct low energy parameters. Two different patches in the
figure correspond to $s_3>0, \cos 2\theta_1<0$ and $s_3<0, \cos
2\theta_1>0$ both of which can lead to the correct solution as argued
above.

In Fig.(2), we display output values of the solar scale $\solm$ and
$m_{ee}$ consistent with the random variation of the input parameters. The
allowed points span the entire range in $\solm$ and thus the two sub
regions corresponding to the KamLand results are easily obtained. The
predicted values of $m_{ee}$ are quite restricted. Typical lower bound in
case of the SM is around $0.05-0.1$ and most points crowd 
in the range $0.4-1.0 \eV$ in case of the SM. Typical range
preferred by MSSM is $m_{ee}\sim 0.2-0.4 \eV$.

We show the  values of $|U_{e3}|$ realized in the random analysis in Fig.(3) as
a function of
$m_{ee}$. In large number of cases, $|U_{e3}|$ is seen to lie
close to the experimental limit in case of the SM.
The MSSM also predicts larger values but allows smaller values $|U_{e3}|\sim
.02$ also.

Fig.(4) is prediction for the mass $m$ which corresponds to the electron
neutrino mass probed in the tritium beta decay. One sees a clear
preference for $m\sim 0.7-2 \eV$. In fact $3\sigma$ lower bound on $m$ 
following from approximate eq.(\ref{d21}) and realized in the figure is $m>0.7$
eV. MSSM also prefers similar values but it can still
allow $m$ as low as $\sim 0.1 \eV$ due to the presence of an additional
parameter $\tan\b$.

We had initially chosen 100,000 random points in both cases but the
allowed set of  input points is much larger in case of the SM.
This clearly shows that radiative corrections in SM can more easily
reproduce the low energy neutrino spectrum than in MSSM. This was to be
expected since when $s_3$ is zero, MSSM cannot lead to the LMA solution at
all \cite{pdp2}.  Introduction of $s_3$ now allows MSSM but the allowed
values of $s_3$ are constrained by the observed bound on $U_{e3}$ and as a
result
one gets correct solution in case of MSSM for much smaller number of input
points. In contrast, the SM can reproduce correct spectrum even for
$s_3=0$ as clearly seen in Fig.(1).

We assumed three neutrinos to be almost degenerate in the above analysis.
Alternative possibility corresponds to the inverted mass hierarchy 
in which only the
solar pair has non-zero mass mass $m$, the third mass $m'$ being zero
or
much smaller. It was argued \cite{pdp1,pdp2} that in all these models the
weak radiative corrections are unable to
give the LMA solution if $s_3=0$. This conclusion arises because the mass $m$
in this case is
required to be close to the atmospheric scale. The resulting   
$\solm$ is at least an order of magnitude smaller than required by LMA in case 
of the SM. MSSM has additional parameter $tan \b$ which
can overcome this suppression but  $tan^2\sola$ is predicted to be
greater than 1 when $s_3=0$. Non-zero $s_3$ changes this conclusion.
As already discussed, one can obtain $\tan^2\sola<1$ even in case of MSSM
if $s_3\cos 2 \theta_1<0$. In this case one can get a $\solm$
corresponding to the LMA solution even for $m^2\sim \atm$ by choosing
a large $\tan\b$. This is demonstrated in Fig. (5) which shows
allowed values of $|s_3|$ and $m$ resulting from the random variations
of input parameters.
Unlike in earlier figures, the mass $m$ is now varied only near the
atmospheric range specifically in the interval $0.01-0.1$ eV and $\cos \b$
is varied in the range $0.001-0.2$ since only these ranges are expected to
give the correct $\solm$ and $\atm$. As seen in the figure, we do get the
correct solutions although for limited 
number of output values starting with 100,000 input points as before.

\noindent {\it Summary and implications for models:}
We have numerically investigated consequences of having vanishing $\solm$
at a high scale. This assumption can account for the smallness of
$\solm$ after inclusion of the weak radiative corrections. The predicted
value for the solar scale is linked to observable low scale
parameters. This results in stringent predictions which can be tested.

Basic assumption of our analysis is two degenerate neutrinos.
This can be realized in number of ways and there are numerous models 
which predict two \cite{emtsym} or all three \cite{deg} neutrinos to be
degenerate.   
If the third neutrino has comparable mass then one gets the almost
degenerate scenario. Precise information on the common mass of these
degenerate neutrinos has recently been provided by the data
on microwave anisotropy \cite{wmap}. In combination with the information 
on the galactic structures, this data imply a bound 
of $m <0.23 eV$ at $95\% CL$  \cite{wmap,mya}. As our
Fig.(4) shows,
this mass is required to be in the range
(0.7-2 ) eV in case of SM but it could be smaller for MSSM. Thus the SM 
radiative corrections and degenerate spectrum cannot account for the LMA
solution at $95\% CL$. The lower bound is also  inconsistent with the 
$99\% CL$ limit, $m<0.3 \eV$  following from analysis by Giunti in
\cite{mya}.  

The standard model also fails in generating correct
$\solm$
if neutrinos have inverted mass hierarchy.
It is possible to obtain the LMA  solution in case of the MSSM
which allows 
degenerate mass in the range $~ 0.1-2 \eV$ range. Likewise, one can also
obtain the LMA solution in case of MSSM and the inverted mass
hierarchy, see Fig.(5).
The LMA solution and value $m<0.23 \eV$ occurs in both these cases only for
very limited range of parameters.

Another related but testable prediction of the scheme is $m_{ee}$ probed
by
$0\nu\b\b$
decay results. The present experimental limit on this scale is uncertain
due to the unknown nuclear matrix element. The present bound is
\cite{limit} $m_{ee}<0.4 h \eV$ at 95\% CL. The $h\sim 0.6-2.8$
parameterizes\footnote{See, Feruglio {\it et al} in \cite{mainz}.}
the uncertainty in nuclear matrix element. The present scenario
is consistent with this limit (see Figs. 2 and 3) but it can be
constrained with improvement on our knowledge of the solar parameters,
$m_{ee}$ and $U_{e3}$.

\newpage
\begin{figure}[h]
\centerline{\psfig{figure=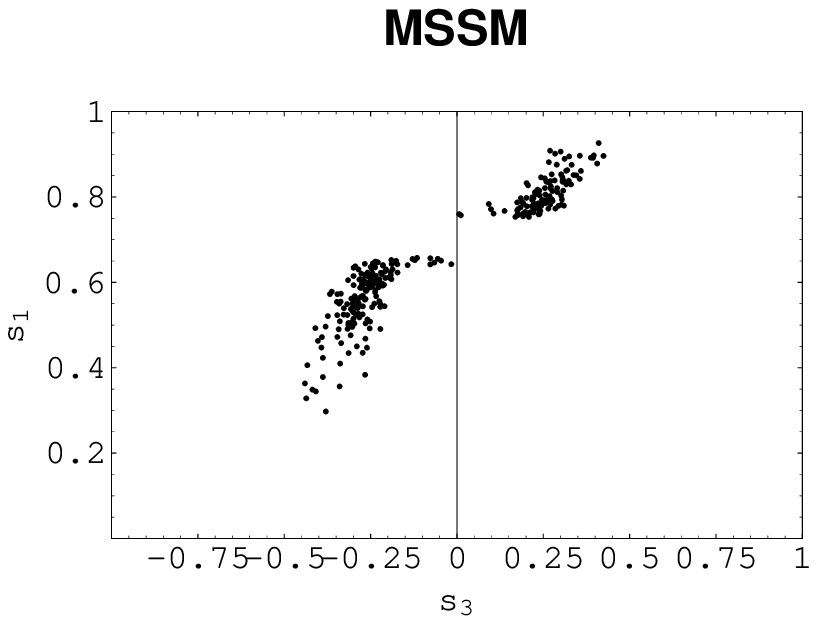,height=8cm,width=11cm}}
\end{figure}
\vskip 1.5truecm
\begin{figure}[h]
\centerline{\psfig{figure=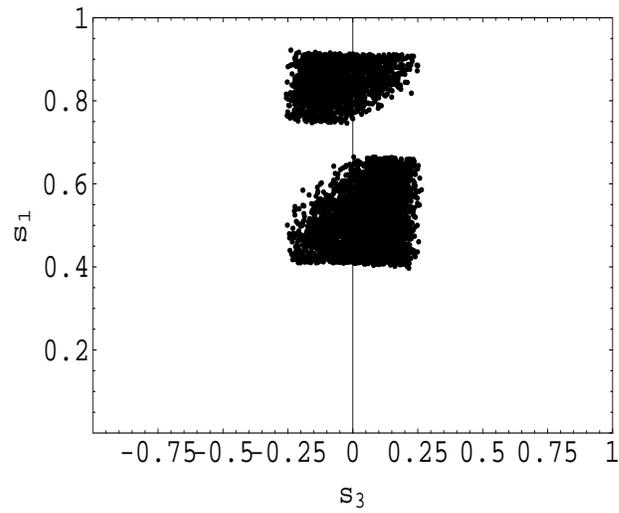,height=8cm,width=10cm}}
\caption{Allowed values of the high scale mixing elements $s_1$ and $s_3$
consistent with the known neutrino
oscillation constraints
in case of the the MSSM and SM. The other input variables are varied
randomly as described in the text.}
\end{figure}
\newpage
\begin{figure}[h]
\centerline{\psfig{figure=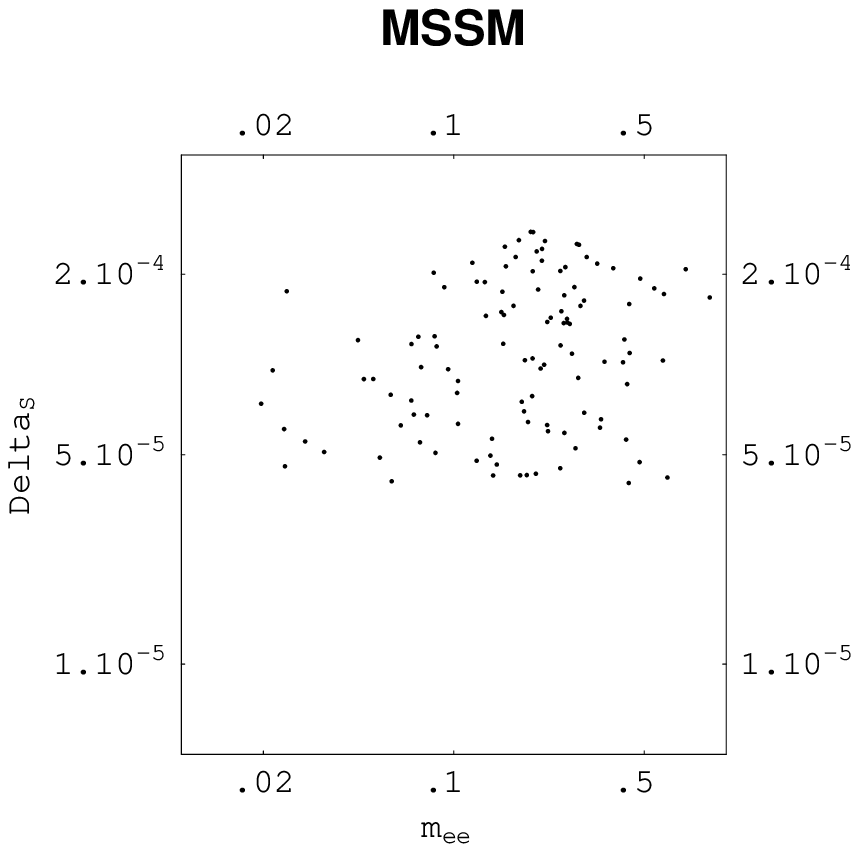,height=7cm,width=11cm}}
\end{figure}
\vskip 1.0truecm
\begin{figure}[h]
\centerline{\psfig{figure=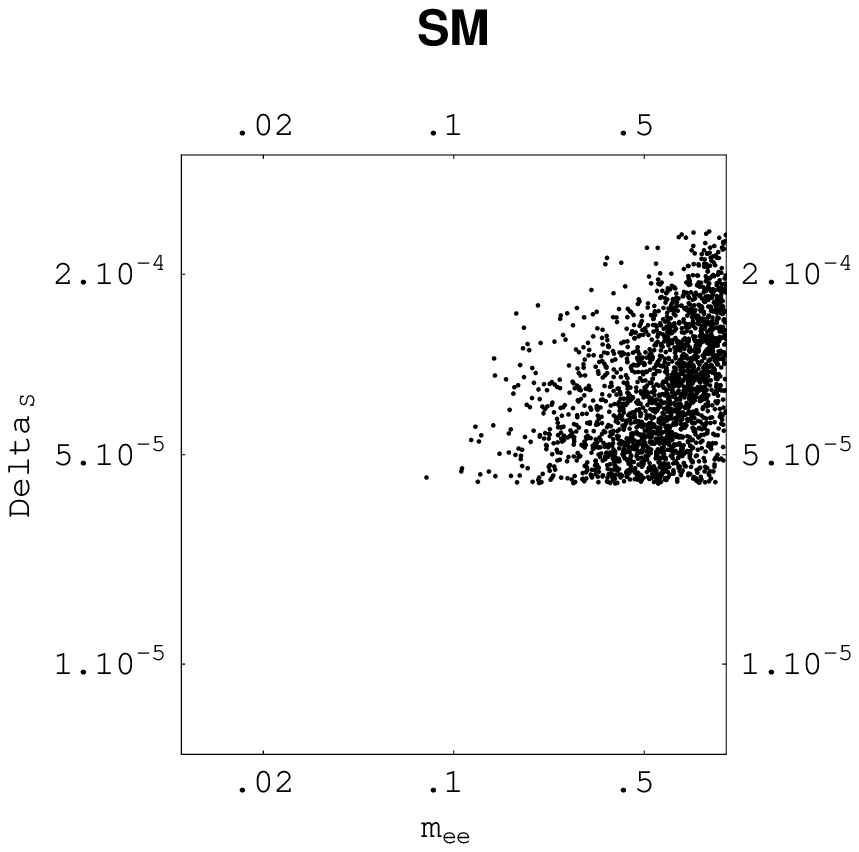,height=7cm,width=11cm}}
\caption{ The predicted correlation between the $0\nu\b\b$ decay parameter 
$m_{ee}$ (in eV) and the solar scale $\solm$ (in $\eV^2$)
resulting after the random variations in input parameters as described in
the text. The upper (lower) figure is for MSSM (SM).}
\end{figure}
\newpage
\begin{figure}[h]
\centerline{\psfig{figure=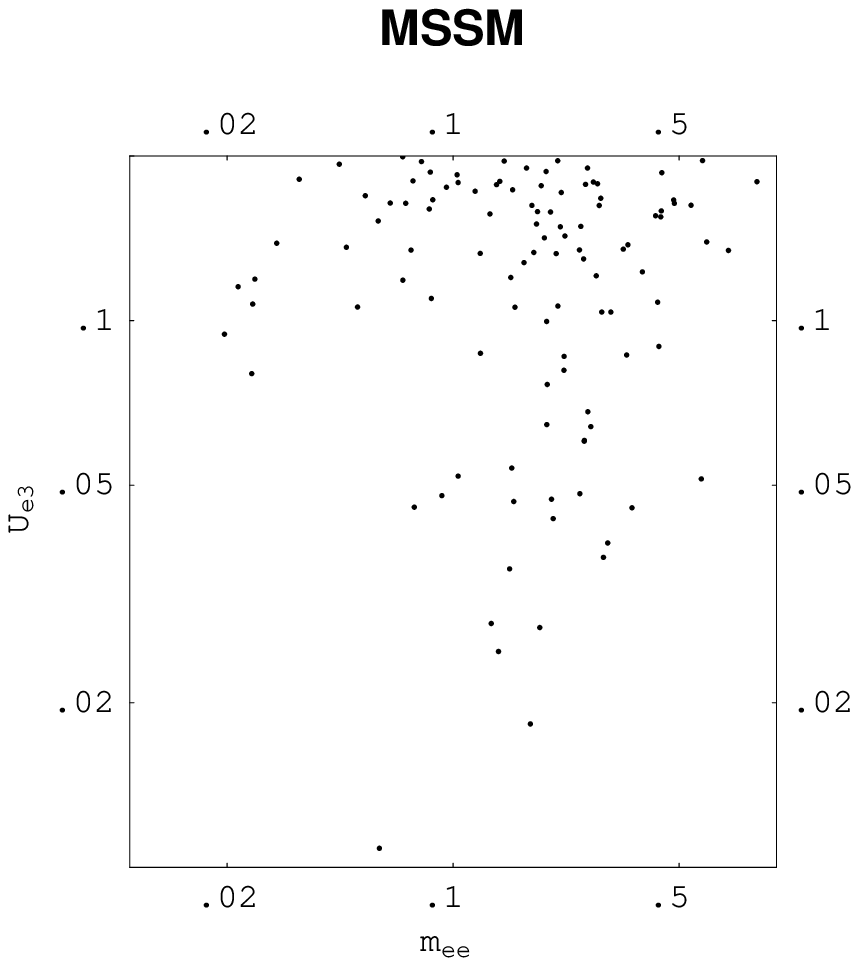,height=8cm,width=11cm}}
\end{figure}
\vskip 1.5truecm
\begin{figure}[h]
\centerline{\psfig{figure=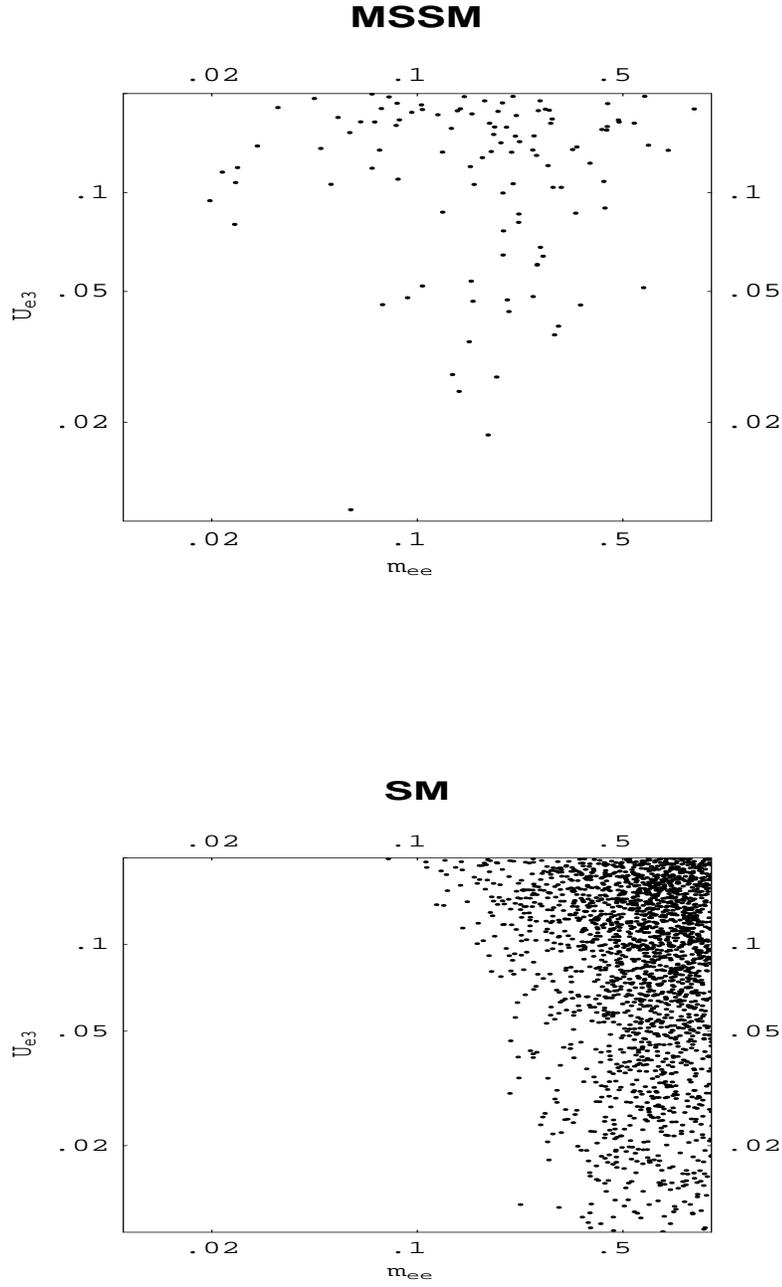,height=7cm,width=11cm}}
\caption{The predicted correlation between the $0\nu\b\b$ decay parameter 
$m_{ee}$ (in eV) and $|U_{e3}|$ 
resulting after the random variations in input parameters as described in
the text. The upper (lower) figure is for MSSM (SM).}
\end{figure}
\newpage
\begin{figure}[h]
\centerline{\psfig{figure=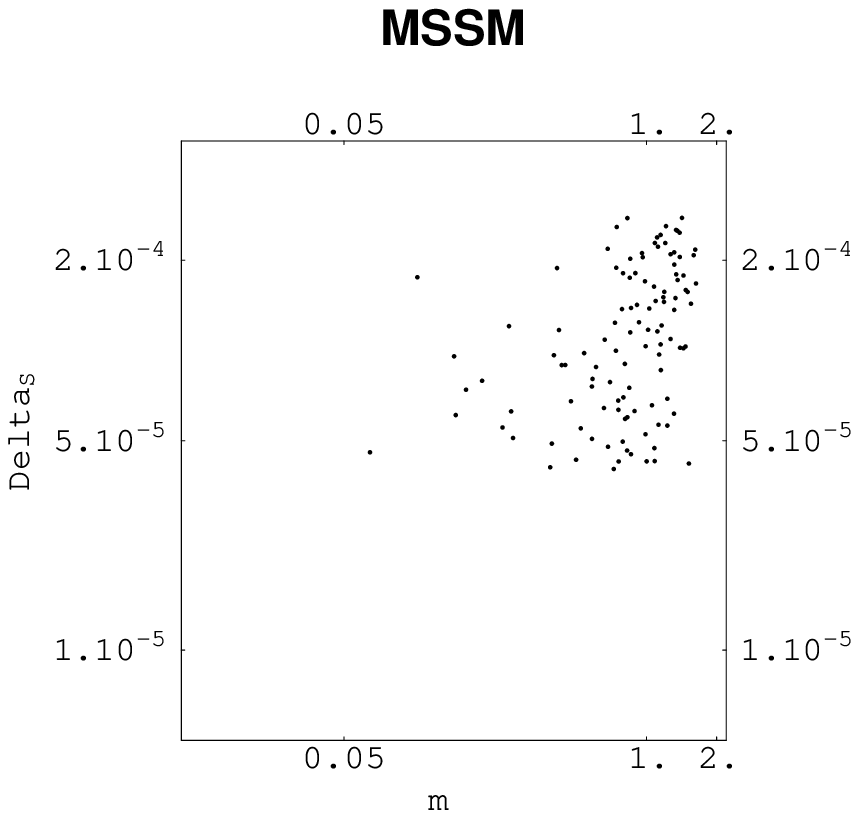,height=7cm,width=11cm}}
\end{figure}
\vskip 1.truecm
\begin{figure}[h]
\centerline{\psfig{figure=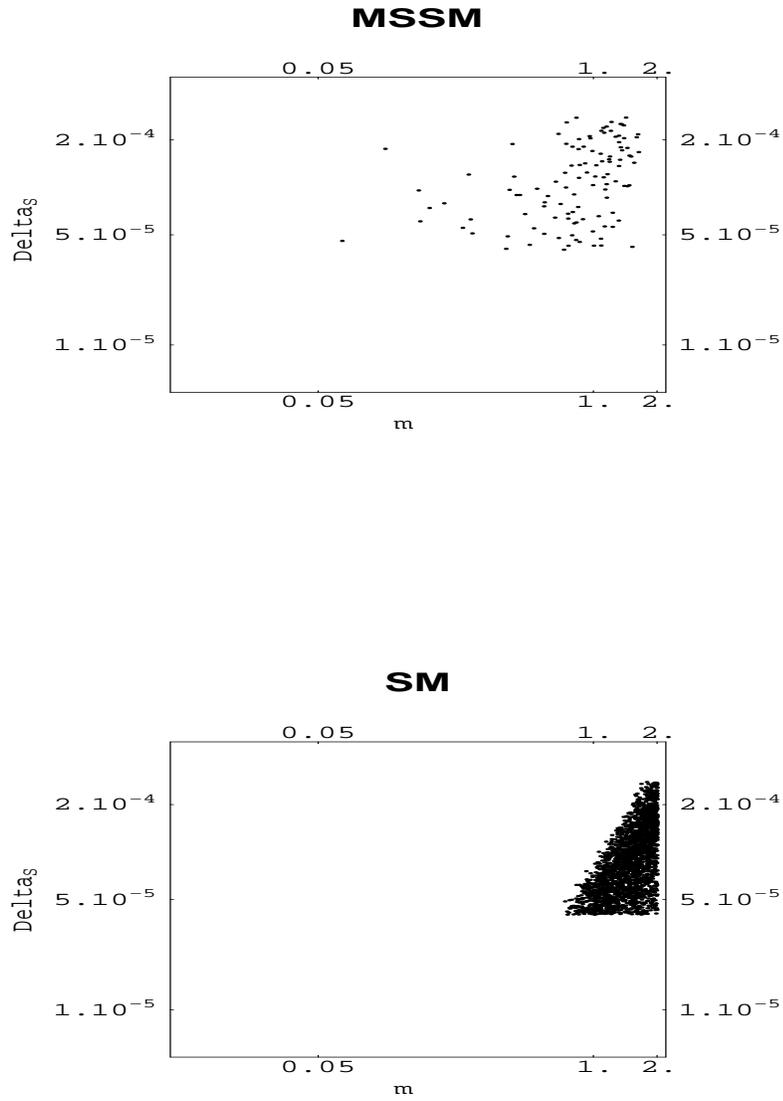,height=7cm,width=11cm}}
\caption{The predicted correlation between the absolute neutrino mass $m$
(in eV) and the solar scale $\solm$ (in $\eV^2$)
resulting after the random variations in input parameters as described in
the text. The upper (lower) figure is for MSSM (SM).}
\end{figure}
\begin{figure}[h]
\centerline{\psfig{figure=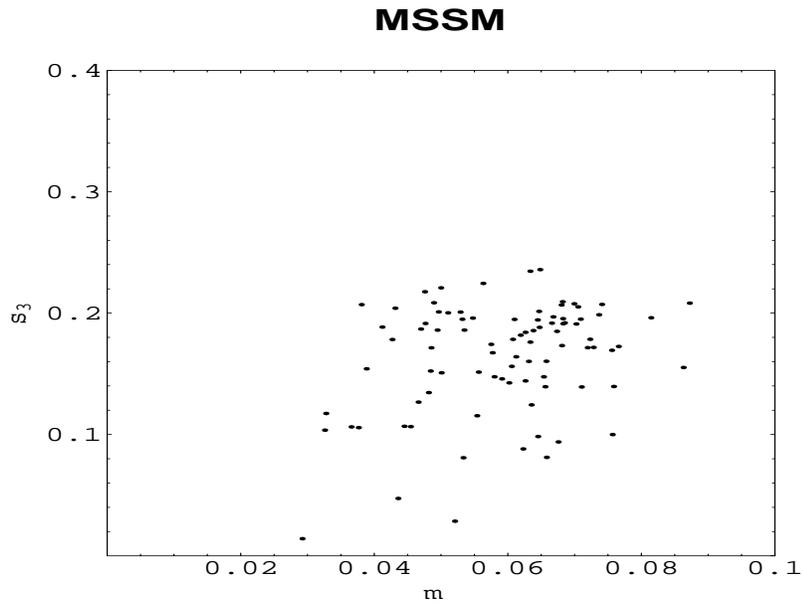,height=8cm,width=11cm}}
\caption{The predicted correlation between the neutrino mass $m$
(in eV) and the absolute value of $s_3$ in inverted hierarchy
models. Input parameters
are randomly varied as described in
the text.}
\end{figure}
\end{document}